\documentclass[twocolumn,english]{IEEEtran}
\usepackage[T1]{fontenc}
\usepackage[latin9]{inputenc}
\usepackage{graphicx}

\makeatletter

\providecommand{\tabularnewline}{\\}

\usepackage{babel}
\makeatother
\usepackage{babel}

\begin{document}

\title{Crowdsourcing Predictors of Behavioral Outcomes}

\author{Josh C.~Bongard, \emph{Member, IEEE}, Paul D.~H.~Hines, \emph{Member,
IEEE}, Dylan Conger, Peter Hurd, and Zhenyu Lu %
\thanks{The authors are with the College of Engineering and Mathematical Sciences,
University of Vermont, Burlington, VT USA (e-mail: \mbox{jbongard}@uvm.edu,
paul.hines@uvm.edu). 

This work was supported in part by the UVM Complex Systems Center, through NASA Grant \#NNX09AJ18G. D. Conger was supported by the McNair Scholars Program.

This paper has been accepted for publication in a future edition of the {\it IEEE Transactions on Systems, Man, and Cybernetics}. Updated March 8, 2012.
}}
\maketitle
\begin{abstract}
Generating models from large data sets---and determining which subsets of data to mine---is becoming increasingly automated.
However choosing what data to collect in the first place requires human intuition or experience, usually supplied by a domain expert.
This paper describes a new approach to machine science which demonstrates for the first time that non-domain experts can collectively formulate features, and provide values for those features such that they are predictive of some behavioral outcome of interest.
This was accomplished by building a web platform in which human groups interact to both respond to questions likely to help predict a behavioral outcome and pose new questions to their peers.
This results in a dynamically-growing online survey, but the result of this cooperative behavior also leads to models that can predict user's outcomes based on their responses to the user-generated survey questions.
Here we describe two web-based experiments that instantiate this approach: the first site led to models that can predict users' monthly electric energy consumption; the other led to models that can predict users' body mass index.
As exponential increases in content are often observed in successful online collaborative communities, the proposed methodology may, in the future, lead to similar exponential rises in discovery and insight into the causal factors of behavioral outcomes.
\end{abstract}

\section{Introduction}

There are many problems in which one seeks to develop predictive models
to map between a set of predictor variables and an outcome. Statistical
tools such as multiple regression or neural networks provide mature
methods for computing model parameters when the set of predictive
covariates and the model structure are pre-specified. Furthermore,
recent research is providing new tools for inferring the structural
form of non-linear predictive models, given good input and output
data \cite{Bongard:2007}. However, the task of choosing which potentially
predictive variables to study is largely a qualitative task that requires
substantial domain expertise. For example, a survey designer must
have domain expertise to choose questions that will identify predictive
covariates. An engineer must develop substantial familiarity with
a design in order to determine which variables can be systematically
adjusted in order to optimize performance. 

The need for the involvement of domain experts can become a bottleneck to new insights.
However, if the wisdom of crowds could be harnessed to produce insight into difficult problems, one might see exponential rises in the discovery of the causal factors of behavioral outcomes, mirroring the exponential growth on other online collaborative communities.
Thus, the goal of this research was to test an alternative approach to modeling in which the wisdom of crowds is harnessed to both propose potentially predictive variables to study by asking questions, and respond to those questions, in order to develop a predictive model.

\subsection*{Machine science}

Machine science \cite{Evans:2010} is a growing trend that attempts
to automate as many aspects of the scientific method as possible.
Automated generation of models from data has a long history, but recently
robot scientists have been demonstrated that can physically carry
out experiments \cite{King:04,King:09} as well as algorithms that
cycle through hypothesis generation, experimental design, experiment
execution, and hypothesis refutation \cite{Bongard:2006,Bongard:2007}.
However one aspect of the scientific method that has not yet yielded
to automation is the selection of variables for which data should
be collected to evaluate hypotheses. In the case of a prediction problem,
machine science is not yet able to select the independent variables
that might predict an outcome of interest, and for which data collection
is required.

This paper introduces, for the first time, a method by which non domain
experts can be motivated to formulate independent variables as well
as populate enough of these variables for successful modeling. In
short, this is accomplished as follows. Users arrive at a website
in which a behavioral outcome (such as household electricity usage
or body mass index, BMI) is to be modeled. Users provide their own
outcome (such as their own BMI) and then answer questions that may
be predictive of that outcome (such as `how often per week do you
exercise'). Periodically, models are constructed against the growing
data set that predict each user's behavioral outcome. Users may also
pose their own questions that, when answered by other users, become
new independent variables in the modeling process. In essence, the
task of discovering and populating predictive independent variables
is outsourced to the user community.

\subsection*{Crowdsourcing}

The rapid growth in user-generated content on the
Internet is an example of how bottom-up interactions can, under some
circumstances, effectively solve problems that previously required
explicit management by teams of experts \cite{Giles:2005}. Harnessing
the experience and effort of large numbers of individuals is frequently
known as {}``crowdsourcing\textquotedblright{} and has been used
effectively in a number of research and commercial applications \cite{Brabham:2008}.
For an example of how crowdsourcing can be useful, consider Amazon's
Mechanical Turk. In this crowdsourcing tool a human describes a {}``Human
Intelligence Task\textquotedblright{} such as characterizing data
\cite{Sorokin:2008}, transcribing spoken language \cite{Marge:2010},
or creating data visualizations \cite{Kong:2010}. By involving large
groups of humans in many locations it is possible to complete tasks
that are difficult to accomplish with computers alone, and would be
prohibitively expensive to accomplish through traditional expert-driven
processes \cite{Kittur:2008}.

Although arguably not strictly a crowdsourced system, the rapid rise of Wikipedia illustrates how online collaboration can be used to solve difficult problems (the creation of an encyclopedia) without
financial incentives. Ref. \cite{Wightman:2010} reviews several crowdsourcing
tools and argues that direct motivation tasks (tasks in which users
are motivated to perform the task because they find it useful, rather
than for financial motivation) can produce results that are superior
to financially motivated tasks. Similarly, ref. \cite{Wightman:2010}
reports that competition is useful in improving performance on a task
with either direct or indirect motivation.
This paper reports on two tasks with direct motivation:  
for the household energy usage task, users are motivated to understand their home energy usage as a means to improve their energy efficiency;
for the body mass index task, users are motivated to understand their lifestyle choices in order to approach a healthy body weight.
Both instantiations include an element of competition by allowing participants to see how they compare  with other participants and by ranking the predictive quality of questions that participants provide.

There is substantial evidence in the literature and commercial applications
that laypersons are more willing to respond to surveys and queries
from peers than from authority figures or organizations. For example
within the largest online collaborative project, Wikipedia, article writers
often broadcast a call for specialists to fill in details on a particular
article. The response rates to such peer-generated requests are enormous,
and have led to the overwhelming success of this particular project.
In the open source community, open source software that crashes automatically
generates a debug request from the user. Microsoft adopted this practice
but has found that users tend not to respond to these requests, while
responses to open source crashes are substantially higher \cite{Fitzgerald:2006,Howe:2009}.
Medpedia, a Wikipedia-styled crowdsourced system, increasingly hosts
queries from users as to what combinations of medications work well
for users on similar medication cocktails. The combinatorial explosion
of such cocktails is making it increasingly difficult for health providers
to locate such similar patients for comparison without recourse to
these online tools.

Collaborative systems are generally more scalable than top-down systems.
Wikipedia is now orders-of-magnitude larger than Encyclopedia Britannica.
The climateprediction.net project has produced over 124 million hours
of climate simulation, which compares favorably with the amount of
simulation time produced by supercomputer simulations. User-generated
news content sites often host as many or more readers than conventional
news outlets \cite{Thurman:2008}. Finally, many of the most recent
and most successful crowdsourced systems derive their success from
their viral \cite{DiBona:2005,Leskovec:2007} nature: they are designed
such that selective forces exerted by users lead to an exponential
increase in content, automated elimination of inferior content, and
automated propagation of quality content \cite{Lerman:2006}.

Citizen science \cite{Anderson:2002,Cohn:2008,Silvertown:2009} platforms
are a class of crowdsourcing systems that include non-scientists in
the scientific process. The hope is that participants in such systems
are motivated ideologically, as their contributions forward what they
perceive as a worthy cause. In most citizen science platforms user
contributions are `passive': they contribute computational but not
cognitive resources \cite{Anderson:2002,Beberg:2009}. Some platforms
allow users to actively participate by searching for items of interest
\cite{Lintott:2008} or solve problems through a game interface \cite{Cooper:2010}.
The system proposed here falls into this latter category: users are
challenged to pose new questions that, when answered by enough of
their peers, can be used by a model to predict the outcome of interest.

Finally, problem solving through crowdsourcing can produce novel,
creative solutions that are substantially different from those produced
by experts. An iterative, crowdsourced poem translation task produced
translations that were both surprising and preferable to expert translations
\cite{Kittur:2010}. We conjecture that crowdsourcing the selection
of predictive variables can reveal creative, unexpected predictors
of behavioral outcomes. For problems in which behavioral change is
desirable (such as is the case with obesity or energy efficiency),
identifying new, unexpected predictors of the outcome may be useful
in identifying relatively easy ways for individuals to change their
outcomes.

\section{Methodology}

\begin{figure}[!t] 
 \centering \includegraphics[width=1.0\columnwidth]{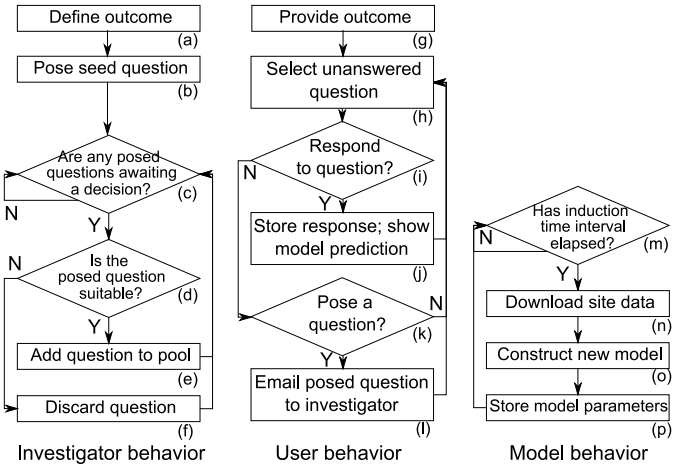}
\caption{ \label{FigOverview} \textbf{Overview of the system.}
The investigator (\textbf{a-f}) is responsible for initially creating the web platform, and seeding it with a starting question. Then, as the experiment runs they filter new survey questions generated by the users. Users (\textbf{g-l}) may elect to answer as-yet unanswered survey questions or pose some of their own. The modeling engine (\textbf{m-p}) continually generates predictive models using the survey questions as candidate predictors of the outcome and users' responses as the training data.}
\end{figure}

The system described here wraps a human behavior modeling paradigm
in cyberinfrastructure such that: (1) the investigator defines some
human behavior-based outcome that is to be modeled; (2) data is collected
from human volunteers; (3) models are continually generated automatically;
and (4) the volunteers are motivated to propose new independent variables.
Fig. \ref{FigOverview} illustrates how the investigator, participant
group and modeling engine work together to produce predictive models
of the outcome of interest. The investigator begins by constructing
a web site and defining the human behavior outcome to be modeled (Fig.
\ref{FigOverview}a). In this paper a financial and health outcome
were investigated: the monthly electric energy consumption of an individual
homeowner (Sect. \ref{sectEnergy}), and their body mass index (Sect.
\ref{sectBMI}). The investigator then initializes the site by seeding
it with a small set (one or two) of questions known to correlate with
the outcome of interest (Fig. \ref{FigOverview}b). For example, based
on the suspected link between fast food consumption and obesity \cite{Bowman:2004,Currie:2010},
we seeded the BMI website with the question {}``\textit{How many
times a week do you eat fast food?}''

\begin{figure*}[!t] 
 \centering \includegraphics[width=1\textwidth]{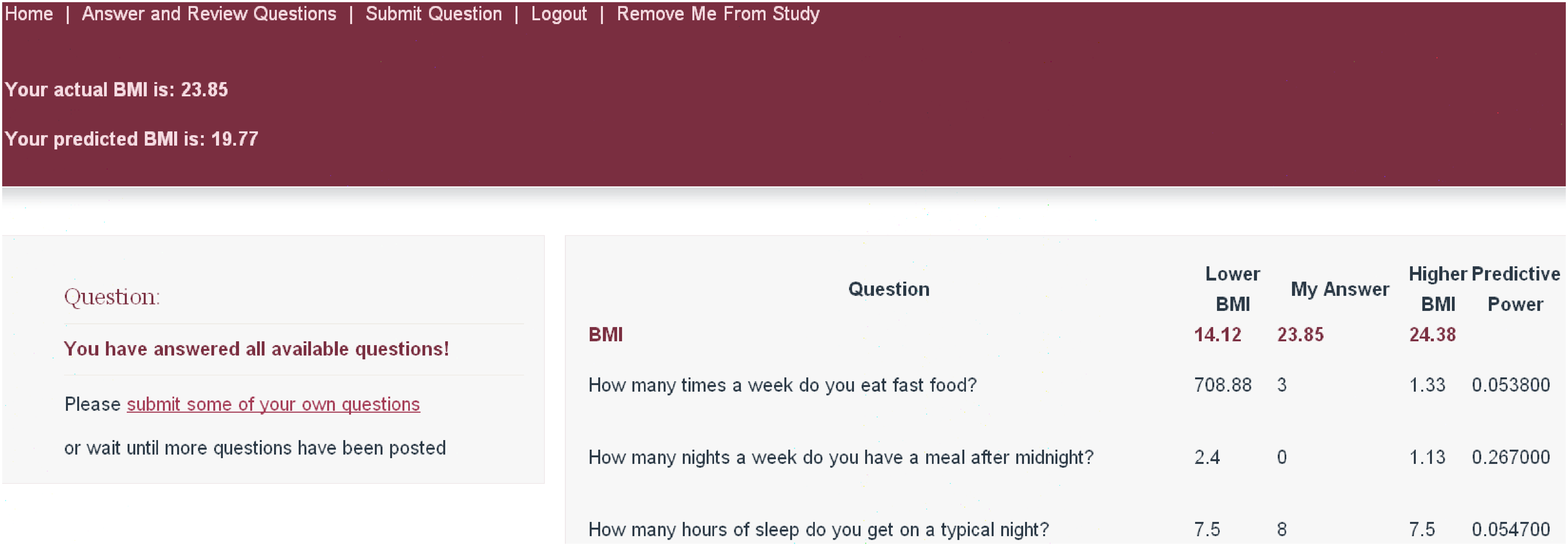}
\caption{\label{figScreenBMI}Screenshot from the Body Mass Index (BMI) website
as seen by a user who has responded to all of the available questions.
The user has the option to change their response to a previous question,
pose a new question, or remove themselves automatically from the study.}
\end{figure*}

Users who visit the site first provide their individual value for
the outcome of interest, such as their own BMI (Fig. \ref{FigOverview}g).
Users may then respond to questions found on the site (Fig. \ref{FigOverview}h,i,j).
Their answers are stored in a common data set and made available to
the modeling engine. Periodically the modeling engine wakes up (Fig.
\ref{FigOverview}m) and constructs a matrix $\mathbf{A}\in\Re^{n\times k}$
and outcome vector $\mathbf{b}$ of length $n$ from the collective
responses of $n$ users to $k$ questions (Fig. \ref{FigOverview}n).
Each element $a_{ij}$ in $\mathbf{A}$ indicates the response of
user $i$ to question $j$, and each element $b_{i}$ in $\mathbf{b}$
indicates the outcome of interest as entered by user $i$. In the
work reported here linear regression was used to construct models
of the outcome (Fig. \ref{FigOverview}o), but any model form could
be employed. The modeling process outputs a vector $\mathbf{c}$ of
length $k+1$ that contains the model parameters. It also outputs
a vector $\mathbf{d}$ of length $k$ that stores the predictive power
of each question: $d_{j}$ stores the $r^{2}$ value obtained by regressing
only on column $j$ of $\mathbf{A}$ against the response vector $\mathbf{b}$.
These two outputs are then placed in the data store (Fig. \ref{FigOverview}p).

At any time a user may elect to pose a question of their own devising
(Fig. \ref{FigOverview}k,l). Users could pose questions that required
a yes/no response, a five-level Likert rating, or a number. Users
were not constrained in what kinds of questions to pose. However,
once posed, the question was filtered by the investigator as to its
suitability (Fig. \ref{FigOverview}d). A question was deemed unsuitable
if any of the following conditions were met: (1) the question revealed
the identity of its author (e.g. {}``\textit{Hi, I am John Doe. I
would like to know if...}'') thereby contravening the Institutional
Review Board approval for these experiments; (2) the question contained
profanity or hateful text; (3) the question was inappropriately correlated
with the outcome (e.g. {}``\textit{What is your BMI?}''). If the
question was deemed suitable it was added to the pool of questions
available on the site (Fig. \ref{FigOverview}e); otherwise the question
was discarded (Fig. \ref{FigOverview}f).

Each time a user responded to a question, they were shown a new, unanswered
question as well as additional data devised to maintain interest in
the site and increase their participation in the experiment. Once
a user had answered all available questions, they were shown a listing
of the questions, their responses, and contextual information to indicate
how their responses compared to those of their peers. Fig. \ref{figScreenBMI}
shows the listing that was shown to those users who participated in
the BMI site; the individual elements are explained in more detail
in Sect. \ref{sectBMI}.

The most important datum shown to each user after responding to each
question was the value of their actual outcome as they entered it
($b_{i}$) as well as their outcome as predicted by the current model
($\hat{b}_{i}$). Fig. \ref{figScreenBMI} illustrates that visitors
to the BMI site were shown their actual BMI (as entered by them) and
their predicted BMI. The models were able to predict each user's outcome
before they had responded to every question by substituting in missing
values. Thus after each response from a user

\begin{equation}
\hat{b}_{i}=c_{0}+c_{1}a_{i1}+c_{2}a_{i2}+...+c_{k}a_{ik}+\epsilon_{i}\label{eq:Regression-model}
\end{equation}
 where $a_{ij}=0$ if user $i$ has not yet responded to question
$j$ and $a_{ij}$ is set to the user's response otherwise.

\section{Energy efficiency instantiation and results\label{sectEnergy}}

In the first instantiation of this concept, we developed a web-based
social network to model residential electric energy consumption. Because
of policy efforts to increase energy efficiency, many are working
to provide consumers with better information about their energy consumption.
Research on consumer perception of energy efficiency indicates that
electricity customers often misjudge the relative importance of various
activities and devices to reducing energy consumption \cite{Attaria:2010}.
To provide customers with better information, numerous expert-driven
web-based tools have been deployed \cite{ms-hohm,Mills:2008,Allcott:2011}.
In some cases these tools use social pressure as a means of improving
energy efficiency \cite{Petersen:2007,Kaufman:2009}, however the
feedback provided to customers typically comes from a central authority
(i.e., top-down feedback) and research on risk perception \cite{Slovic:1999}
indicates that the public is often skeptical of expert opinions. A
recent industry study \cite{Guthridge:2010} indicates that customers
are notably skeptical of large online service providers (e.g., Google,
Microsoft) and (to a lesser extent) electric utilities as providers
of unbiased information about energy conservation. Therefore, information
generated largely by energy consumers themselves, in a bottom-up fashion,
may have value in terms of motivating energy efficient behavior.

Thus motivated, we designed the {}``EnergyMinder'' website to predict
and provide feedback about monthly household (residential) electricity
consumption. Participants were invited to join the site through notices
in university e-mail networks, a university news letter, and reddit,
a user-generated content news site. The site was launched in July
of 2009, and gradually accumulated a total of 58 registered users
by December of 2009. The site consisted of a simple login page and
five simple, interactive pages. The\emph{ Home Page} (after login)
contained a simple to-do list pointing users to tasks on the site,
such as, enter bill data, answer questions, check their energy efficiency
ranking, etc. The \emph{Energy Input Page} showed a time series trend
of the consumer's monthly electricity consumption and asked the user
to enter the kilowatt hours (kWh) consumed for recent months. This
value became the output variable ($b_{i}$) in the regression model
(Eq. \ref{eq:Regression-model}) for a particular month. The \emph{Ask-A-Question
Page} allowed users to ask questions of the group, such as {}``How
many pets do you have?'' (Question 10, Table \ref{tab:EM-Questions}).
When typing in a new question, users were instructed to specify the
type of answer expected (numeric, yes/no, agree/disagree) and to provide
their own response to the question. The \emph{Answer Page} asked participants
to respond to questions, and provided them with information about
each answered question including the distribution of answers within
the social network. Finally, a \emph{Ranking Page} showed users their
energy consumption, relative to that of others in the group. In addition
the Ranking Page reported the predictive power (the percentage of
explained variance) for each statistically significant question/factor.
This final page was intended to provide information to participants
that might help them in choosing behaviors that would reduce electricity
consumption.

\begin{table*}
\caption{\label{tab:EM-Questions}Questions entered into the EnergyMinder web
site. }
\begin{tabular}{clccccccc}
\hline
 &  &  & \# of  & answers  & { Model 1{**}} &  & { Model 2{**}} & \tabularnewline
 & { Question} & { Type} & answers & { in $G$ } & { $c_{i}$} & { $P$ } & { $c_{i}$} & { $P$}\tabularnewline
\hline
{ 1.} & { What is the square footage of your house?{*}} & { numeric} & { 45} & { 22} & { 0} & { 0.52} & { -} & { -}\tabularnewline
{ 2.} & { How many children do you live with?{*}} & { numeric} & { 41} & { 22} & { 109} & { 0.47} & { -} & { -}\tabularnewline
{ 3.} & { How many adults do you live with?} & { numeric} & { 43} & { 22} & { 303} & { 0.03} & { 297} & {{} 0.01 }\tabularnewline
{ 4.} & { How many south facing windows do you have?} & { numeric} & { 42} & { 22} & { -11} & { 0.77} & { -} & { -}\tabularnewline
{ 5.} & { Do you have an electric clothes dryer?} & { yes/no} & { 35} & { 19} & { 430} & { 0.23} & { 240} & {{} 0.28 }\tabularnewline
{ 6.} & { Do you have an electric water heater?} & { yes/no} & { 33} & { 18} & { -577} & { 0.04} & { -535} & {{} 0.01 }\tabularnewline
{ 7.} & { Do you have gas heating?} & { yes/no} & { 34} & { 18} & { 188} & { 0.44} & { -} & { -}\tabularnewline
{ 8.} & { Do you have geothermal heating?} & { yes/no} & { 16} & { 10} & { -} & { -} & { -} & { -}\tabularnewline
{ 9.} & { How many adults are typically home throughout the day?} & { numeric} & { 17} & { 10} & { -} & { -} & { -} & { -}\tabularnewline
{ 10.} & { How many pets do you have?} & { numeric} & { 15} & { 9} & { -} & { -} & { -} & { -}\tabularnewline
\hline
 & { $r^{2}$ value for predictive models} &  &  &  & { 0.63} &  & { 0.57} & \tabularnewline
\hline
\end{tabular}

{ {*} Questions 1 and 2 were seed questions placed on
the site by the investigators.}{ \par}

{ {*}{*} In Model 1 and Model 2, $c_{i}$ is the parameter
estimate ($\mbox{kWh}\cdot\mbox{month}^{-1}\cdot\mbox{unit}^{-1})$
and $P$ is the significance level of the parameter estimate.}
\end{table*}
\begin{figure}
\includegraphics[width=1\columnwidth]{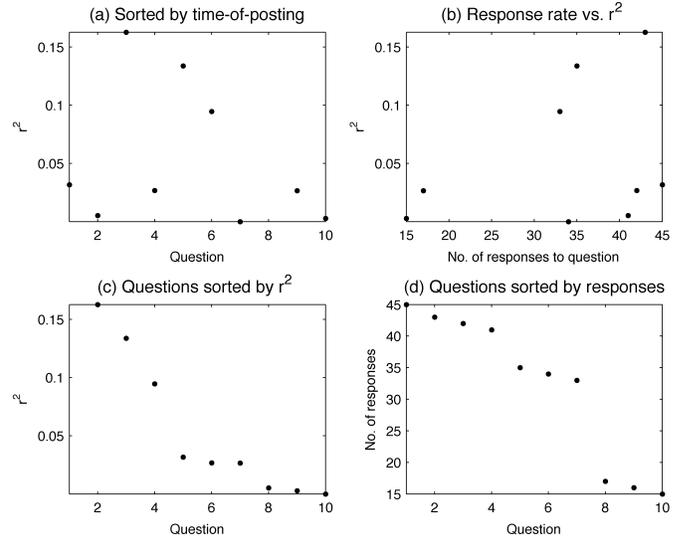}

\caption{\label{fig:e-minder}\textbf{EnergyMinder Question Statistics}. Panel
(\textbf{a}) shows the $r^{2}$ value for each question as numbered in table
\ref{tab:EM-Questions}. (\textbf{b}) shows that there is a mild correlation
between the response rate and the $r^{2}$ values. (\textbf{c}) shows the questions
sorted by their $r^{2}$ value, and (\textbf{d}) shows the number of responses
for each question, sorted by the number of responses.}
\end{figure}

In total the site attracted 58 participants, of whom 46 answered one
or more questions, and 33 (57\%) provided energy consumption data.
Eight new questions were generated by the group, after the seed questions
($Q_{1}$ and $Q_{2}$ in Table \ref{tab:EM-Questions}) were placed
there by the investigators. The fact that only about half of the participants
provided energy data was most likely due to the effort associated
with finding one or more electricity bills and entering data into
the site. This low response rate emphasized that the utility of this
approach depends highly on the ease with which the user can access
the outcome data.

Despite the small sample size, this initial trial resulted in a statistically
significant predictive model, and provided insight into the nature
of the method. Of the 33 participants, 24 provided data for the months
of June, July or August. Because this was the largest period for which
common data were available, the mean outcome for these three months
was used as the outcome variable $b_{i}$. One participant reported
kWh values that were far outside of the mean (46,575 kWh per month)
and one did not answer any questions. These two data sets were discarded
as outliers. The $N=22$ that remained comprised the sample-set used
to produce the results that follow.

Table \ref{tab:EM-Questions} shows results from two predictive models.
Model 1 included all questions that had 18 or more answers ($Q_{1}$-$Q_{7}$).
The total explained variance for Model 1 was $r^{2}=0.63$. Model
1 indicated that the number of adults in the home ($Q_{3}$) significantly
increased monthly electricity consumption ($P<0.05$) and the ownership
of a natural gas hot water heater ($Q_{6}$) significantly decreased
electricity consumption ($P<0.05$). Note that this second result
is not consistent with the fact that owning an electric hot water
heater increases electricity consumption. It appears either that this
correlation was due to chance, or that ownership of a gas hot water
heater correlates to some other factor, such as (for example) home
ownership. Model 2 tested the removal of the least significant predictors,
and included only $Q_{3}$, $Q_{5}$, and $Q_{6}$. Model 2 showed
the same pair of statistically significant predictors ($Q_{3}$ and
$Q_{6}$).

Figure \ref{fig:e-minder} shows the relative predictive power of
the 10 questions. The results show that the most highly correlated
factors ($Q_{3}$, $Q_{5}$, and $Q_{6}$) were posed after the initial
two seed questions (Fig. \ref{fig:e-minder}a) and a weak correlation between
the response rate and the $r^{2}$ values, indicating that more answers
to questions would have likely produced improved results.
Panels (c) and (d) show the distributions of $r^2$ values and the number of responses, to facilitate comparison with the BMI results (Fig. \ref{BMI_Fig3}).

While the small sample size in this study limits the generality of
these results, this initial trial provided useful information about
the crowdsourced modeling approach. Firstly, we found that participants
were reluctant or unable to provide accurate outcome data due to the
challenge of finding one's electric bills. Our second experiment corrects
this problem by focusing on an outcome that is readily accessible
to the general public. Secondly, we found that participants were quite
willing to answer questions posed by others in the group. Questions
1-4 were answered by over 70\% of participants. This indicated that
it is possible to produce user-generated questions and answers, and
that a trial with a larger sample size might provide more valuable
insight. Finally, questions that were posed early in the trial gained
a higher response rate, largely because many users did not return
to the site after one or two visits. This emphasizes the importance
of attracting users back to the site to answer questions in order
to produce a statistically useful model.

\section{Body Mass Index instantiation and results\label{sectBMI}}

In order to test this approach with an outcome that was more readily
available to participants a second website was deployed in which models
attempted to predict the body mass index of each participant. Body
mass index (BMI) is calculated as mass(kg) / (height(m))$^{2}$ and,
although it is known to have several limitations \cite{Romero:2008},
is still the most common measure for determining a patient's level
of obesity. Each user's BMI could readily be calculated as all users
know and are thus able to immediately enter their height and weight.
A second motivator for investigating this behavioral outcome is that
obesity has been cited \cite{Barness:2007} as one of the major global
public health challenges to date, it is known to have myriad causes
\cite{Parsons:1999,Wang:2007}, and people with extreme BMI values
are likely to have intuitions as to why they deviate so far from the
norm.

\begin{table*}
\begin{centering}
\caption{\label{tableBMI}Listing of the 20 most predictive questions from
the BMI site.}

\par\end{centering}

\centering{}%
\begin{tabular}{clcc}
\hline
Index  & Question  & $r^{2}$  & Responses \tabularnewline
\hline
1  & Do you think of yourself as overweight?  & 0.5524  & 43 \tabularnewline
2  & How often do you masturbate a month?  & 0.3887  & 32 \tabularnewline
3  & What percentage of your job involves sitting?  & 0.3369  & 57 \tabularnewline
4  & How many nights a week do you have a meal after midnight?  & 0.2670  & 67 \tabularnewline
5  & You would consider your partner/boyfriend/girlfriend/spouse etc to
be overweight?  & 0.2655  & 24 \tabularnewline
6  & How many, if any, of your parents are obese?  & 0.2311  & 57 \tabularnewline
7  & Are you male?  & 0.2212  & 32 \tabularnewline
8  & I am happy with my life  & 0.2062  & 31 \tabularnewline
9  & How many times do you cook dinner in an average week?  & 0.2005  & 44 \tabularnewline
10  & How many miles do you run a week?  & 0.1865  & 28 \tabularnewline
11  & Do you have a college degree?  & 0.1699  & 12 \tabularnewline
12  & Do you have a Ph.D.  & 0.1699  & 12 \tabularnewline
13  & Would you describe yourself as an emotional person?  & 0.1648  & 30 \tabularnewline
14  & How often do you eat (meals + snacks) during a day  & 0.1491  & 33 \tabularnewline
15  & How many hours do you work per week?  & 0.1478  & 46 \tabularnewline
16  & Do you practice a martial art?  & 0.1450  & 31 \tabularnewline
17  & What is your income?  & 0.1419  & 55 \tabularnewline
18  & I was popular in high school  & 0.1386  & 31 \tabularnewline
19  & Do you ride a bike to work?  & 0.1383  & 64 \tabularnewline
20  & What hour expressed in 1-24 on average do you eat your last meal before
going to bed?  & 0.1364  & 30 \tabularnewline
\hline
\end{tabular}
\end{table*}

Participants arriving for the first time at the BMI site were asked
to enter their height and weight in feet, inches and pounds respectively,
as most of the visitors to the site resided in the U.S. Participants
were then free to respond to and pose new questions.

In order to further motivate the participants, in addition to displaying
their predicted outcome, users were also shown how their responses
compared to two peer groups. For each user the peer groups were constructed
as follows. The first peer group was composed of 10 other users who
had BMI values as close to but below that of the user; the second
group was composed of 10 other users who had BMI values as close to
but above that of the user. If $N<10$ users could be found
the peer group was composed of those $N$ users. The average BMI for
each of the two peer groups, as well as the user's own BMI, were displayed
(see Fig. \ref{figScreenBMI}). Also, the responses to each question,
within each peer group, were averaged and shown alongside the user's
response to that question. Finally, the `predictive power' of each
question was shown. Predictive power was set equal to the $r^{2}$
obtained when the responses to that question alone were regressed
against the outcome.

The peer group data were meant to help users compare how their lifestyle
choices measured up to their most similar peers who were slightly
more healthy than themselves, and slightly less healthy than themselves.
This approach in effect provides individualized suggestions to each
user as to how slight changes in lifestyle choices may lead to improvements
in the health indicator being measured. Presenting the user with the
predictive power of each question was designed to help them learn
what questions tend to be predictive, and thus motivate them to formulate
new or better questions that might be even more predictive. For example
one user posed the question {}``\emph{How many, if any, of your parents
are obese?}''. Another user may realize that the `predictive power'
of this question (which achieved an $r^{2}$ in the actual experiment
of 0.23 and became the sixth-most predictive question out of a total
of 57) may be due to it serving as an indirect measure of the hereditary
component of obesity. This may cause the user to pose a new question
better tailored to eliciting this information, such as {}``\emph{How
many, if any, of your }\textbf{\emph{biological}}\emph{ parents are
obese?}'' (a question of this form was not posed during the actual
experiment).

The BMI site went live at 3:00pm EST on Friday, November 12, 2010,
stayed live for slightly less than a week, and was discontinued at
10:20am EST on Thursday, November 18, 2010. During that time it attracted
64 users who supplied at least one response. Those users proposed
56 questions (in addition to the original seed question), and together
provided 2021 responses to those questions.

\begin{figure}[!t]
 \centering \includegraphics[width=1.0\columnwidth]{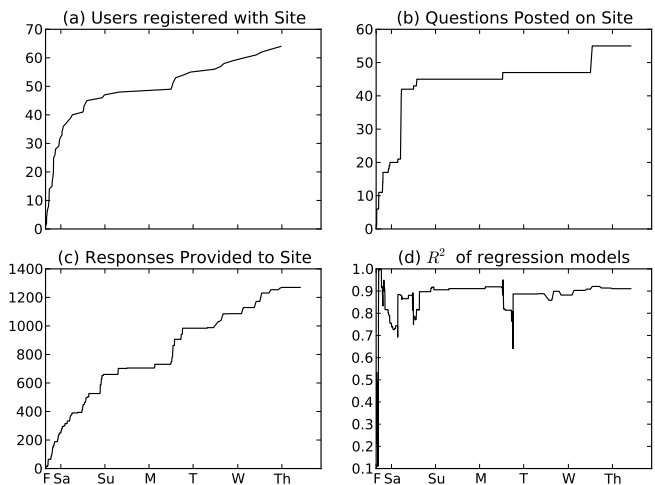}
\caption{ \label{BMI_Fig1} \textbf{User behavior on the BMI site.} The BMI
site was maintained for slightly less than seven days. During that
time it attracted 64 users (\textbf{(a)}) who together posted a total
of 57 questions \textbf{(b)} and 2021 responses to those questions
\textbf{(c)}. Every five minutes a regression model was constructed
against the site's data: The quality of these models are shown as
a function of their $R^{2}$ value \textbf{(f)}. }
\end{figure}

Users were recruited from reddit.com and the social networks of the
principal investigators. Fig. \ref{BMI_Fig1}a shows an initial burst
of new users followed by a plateau during the weekend, and then a
steady rise thereafter until the termination of the experiment. Fig.
\ref{BMI_Fig1}b shows a similar, initially rapid increase in the
number of questions, and no significant increase until one user submits
8 new questions on day 6. Fig. \ref{BMI_Fig1}c shows a relatively
steady rise in the number of responses collected per day. This can
be explained by the fact that although fewer users visit the site
from the third day onward, there are more questions available when
they do and thus, on average, more responses are supplied by later
users than earlier users.

\begin{figure}[!t]
 \centering \includegraphics[width=1.0\columnwidth]{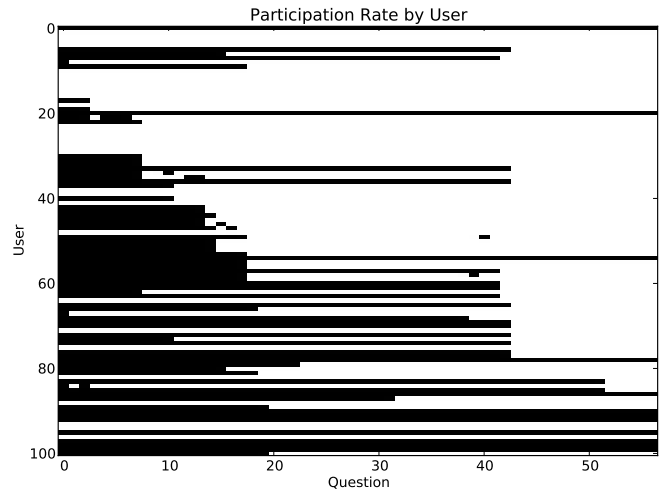}
\caption{ \label{BMI_Fig2} \textbf{Participation Rate by User of the BMI site.}
Each row corresponds to a user of the BMI site, sorted by time of registration. 
Each column corresponds to one of the questions, sorted by time of posting.
A black pixel at row $i$ column $j$ indicates that user $i$ responded to question $j$; a white pixel indicates they did not.}
\end{figure}

This increase is supplemented by a few early users who return to the
site and respond to new questions, as shown in Fig. \ref{BMI_Fig2}.
It shows that of the 100 users who registered, only 57 supplied at
least one response. The triangular form of the matrix is due to the
fact that for the majority of users, they only visited the site once
and answered the questions that were available at that time. This
led to a situation in which questions posted early received disproportionally
more responses than those questions posted later.

For the first several hours of the experiment the modeling engine
(Fig. \ref{FigOverview}m-p) was run once every minute. At 5:30pm
on November 12 the modeling engine was set to run once every five
minutes. With the decrease in site activity the modeling engine was
set to run once an hour starting at 2:20pm on November 16 until the
termination of the experiment. Fig. \ref{BMI_Fig1}d reports the $r^{2}$
value of the regression models as the experiment proceeded. During
the first few hours of the experiment when there were more users than
questions (see Fig. \ref{BMI_Fig1}a,b), the early models had an $r^{2}$
near 1.0, suggesting that overfitting of the data was occurring. However
at the termination of the experiment when there were more users (64)
than questions (57)---and many users had not responded to those questions---the
models were still performing well with an $r^{2}$ near 0.9. There
is still a possibility though that the models overfit the data as
the site was not instrumented with the ability to create a testing
set composed of users whose responses were not regressed against.

\begin{figure}[!t]
 \centering \includegraphics[width=1.0\columnwidth]{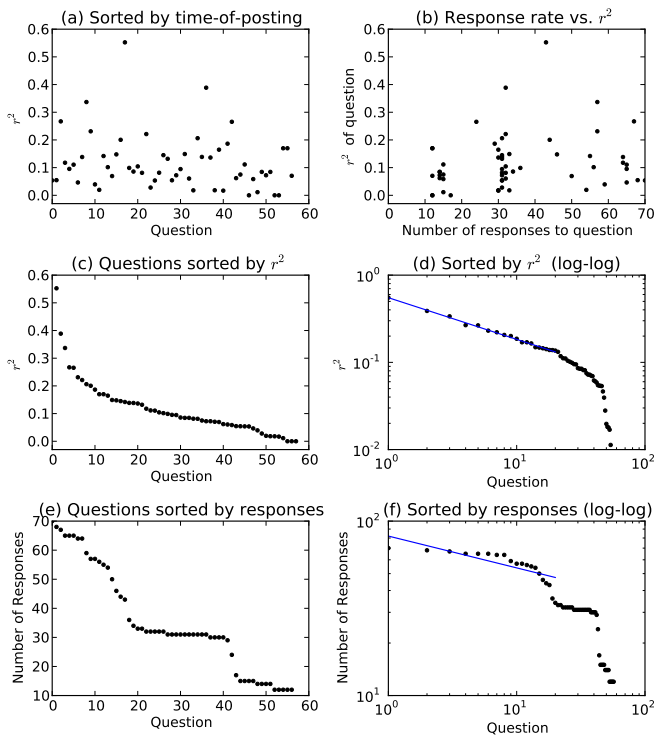}
\caption{\label{BMI_Fig3} \textbf{BMI Question Statistics.}
(\textbf{a,b}): No relationship was found between questions' time of posting, response rate or predictive power.
However a power law relationship was discovered among questions' predictive power (\textbf{c,d}) but not for their response rate (\textbf{e,f}).
}
\end{figure}

Fig. \ref{BMI_Fig3} reports statistics about the user-posed questions.
Fig. \ref{BMI_Fig3}a shows that there is no correlation between when
a question was posed and how predictive it became: the second- and
fifth-most predictive question were posed as the 35th and 42nd question,
respectively. Similarly, Fig. \ref{BMI_Fig3}b reports the lack of
correlation between the number of responses a question receives and
its final predictive power. Although a slight positive correlation
may exist, several of the most predictive questions (including the
second- and fifth-most) received less than half of all possible responses.

Fig. \ref{BMI_Fig3}c reports the questions sorted in order of decreasing
$r^{2}$, and reveals that this distribution has a long tail: a large
number of questions have low, but non-zero $r^{2}$ when regressed
alone against the outcome. This distribution is replotted in Fig.
\ref{BMI_Fig3}c on a log-log scale. Linear regression was performed
on the 20 most predictive questions (indicated by the line), and the resulting fit was found to be highly correlated, with $r^{2}=0.994$. This finding
suggests that a power law relationship exists among these predictive
questions%
\footnote{The close linear fit for these questions does not guarantee that a
power law exists among these questions, however \cite{Clauset:2009}.
Subsequent work and a larger data set will be required to confirm
if power law relationships do indeed exist among user-generated questions
predictive of a behavioral outcome.%
}. It is possible that the power law exists because of an underlying
power law relationship in the number of responses these questions
attracted. However, Fig. \ref{BMI_Fig3}b indicates there is little
or no correlation between the number of responses a question attracts
and its predictive power. Further, Fig. \ref{BMI_Fig3}e reports the
questions sorted by number of responses, and, when plotted on a log-log
scale (Fig. \ref{BMI_Fig3}f) shows that there is no power law ($r^{2}=0.65$)
among the 20 most responded-to questions. This suggests the power
law relationship among the most predictive questions has some other
cause.

Table \ref{tableBMI} reports the 20 most predictive questions, sorted
by decreasing $r^{2}$. The questions span many of the classes of
factors known (or hypothesized) to influence obesity, including demographic
(q. 7 \cite{Boumtje:2005}), social or economic (qs. 5, 11, 12, 15,
17, 18 \cite{Wang:2007}), genetic (q. 6 \cite{Herbert:2006}), psychological
(qs. 1, 8, 13 \cite{Friedman:1995,vandermerwe:2007}) dietary (qs.
4, 9, 14, 20 \cite{Bonow:2003}), and physical activity-related (qs.
2, 3, 10, 16, 19 \cite{Ewing:2008}). This indicates that although
the majority of participants are unlikely to be experts in the domain
of interest, collectively they uncovered many of the classes of known
correlates of obesity, and responded sufficiently honestly so that
these correlates became predictive of BMI on the site. It could be
argued that the most predictive question should not have been accepted
as it is highly likely that it correlates with the outcome: people
who perceive themselves as overweight are likely to be overweight.
However, it is known that for those suffering from body image disorders
the opposite is often the case: those that perceive themselves incorrectly
as overweight can become extremely underweight \cite{Grogan:2008}.
Separating the auto- and anti-correlated components of this broad
question could be accomplished by supplementing it with more targeted
questions (eg., {}``\textit{Do you think you are overweight but everyone
else tells you the opposite?}'').

Despite the lack of filtering on the site there were only a few cases
of clearly dishonest responses. Fig. \ref{figScreenBMI} indicates
that at least one member of this user's peer group answered the fast
food question dishonestly. It is interesting to note that this dishonest
answer (or answers) was supplied for the seed question, and this question---despite
collecting the most responses (70)---had nearly no individual correlation
($r^{2}=0.054$) and thus contributed negligibly to the predictions
of the models. Questions 3, 4, 6, 9, 15, and 20 as shown in Table
\ref{tableBMI} have maximum possible values (qs. 3 max=100; qs. 4
and 9 max=7; qs. 6 max=2; qs. 15 max=168; qs. 20 max=24), and together
collected 301 responses. Of those responses, none were above the maximum
or below the minimum (min=0 for all qs.) indicating that all responses
were not theoretically impossible. This suggests that clear dishonesty
(defined as supplying a response below or above the theoretical minimum
or maximum, respectively) was quite rare for this experiment. Conversely,
unlike the popular yet corrupted seed question, these questions became
significantly predictive as the experiment progressed. Further investigation
into whether or how the rare cases of clear dishonesty (and the possibly
larger amount of hidden dishonesty) affect modeling in such systems
remains to be investigated.

\section{Discussion/conclusions}

This paper introduced a new approach to social science modeling in which
the participants themselves are motivated to uncover the correlates
of some human behavior outcome, such as homeowner electricity usage
or body mass index. In both cases participants successfully uncovered
at least one statistically significant predictor of the outcome variable.
For the body mass index outcome, the participants successfully formulated
many of the correlates known to predict BMI, and provided sufficiently
honest values for those correlates to become predictive during the
experiment.
While, our instantiations focus on energy and BMI, the proposed method is general, and might, as the method improves, be useful to answer many difficult questions regarding why some outcomes are different than others.
For example, future instantiations might provide new insight into difficult questions like:
"Why do grade point averages or test scores differ so greatly among students?",
"Why do certain drugs work with some populations, but not others?",
"Why do some people with similar skills and experience, and doing similar work, earn more than others?"

Despite this initial success, much work remains to be done to improve
the functioning of the system, and to validate its performance. The
first major challenge is that the number of questions approached the
number of participants on the BMI website. This raises the possibility
that the models may have overfit the data as can occur when the number
of observable features approaches the number of observations of those
features. Nevertheless the main goal of this paper was to demonstrate
a system that enables non domain experts to collectively formulate
many of the known (and possibly unknown) predictors of a behavioral
outcome, and that this system is independent of the outcome of interest.
One method to combat overfitting in future instantiations of the method
would be to dynamically filter the number of questions a user may
respond to: as the number of questions approaches the number of users
this filter would be strengthened such that a new user is only exposed
on a small subset of the possible questions.

\subsection{User Fatigue}

Another challenge for this approach is user fatigue: Fig. \ref{BMI_Fig2}
indicates that many of the later users only answered a small fraction
of the available questions. Thus it is imperative that users be presented
with questions that most require additional responses first. This
raises the issue of how to order the presentation of questions. In
the two instantiations presented here, questions were simply presented
to all users in the same order: the order in which they were posted
to the site. It was possible that this ordering could have caused
a `winner take all' problem in that questions that accrue more responses
compared to other questions would achieve a higher predictive power,
and users would thus be attracted to respond to these more predictive
questions more than the less predictive questions. However, the observed
lack of correlation between response rate and predictive power (Fig.
\ref{BMI_Fig3}b) dispelled this concern.

In future instantiations of the method, question ordering will be
approached in a principled way. Instead of training a single model
$m$, an ensemble of methods $m_{1},...,m_{k}$ will be trained on
different subsets of the data \cite{Skurichina:2002,Lu:2009}. Then,
query by committee \cite{Seung:1992} will be employed to determine
question order: The question that induces maximal disagreement among
the $k$ models as to its predictive power will be presented first,
followed by the question that induces the second largest amount of
disagreement, and so on. In this way questions that may be predictive
would be validated more rapidly than if question ordering is fixed,
or random.

\subsection{User Motivation}

Typically, human subjects play a passive role in social science studies, regardless of whether that study is conducted offline (pen-and-paper questionnaire) or online (web-based survey): They contribute responses to survey questions, but play no role in crafting the questions. This work demonstrates that users can also contribute to the hypothesis-generation component of the discovery process: Users can collectively contribute---and populate---predictors of a behavioral outcome.

It has been shown here that users can be motivated to do this without requiring an explicit reward: The subjects were unpaid for both studies. Much work remains to be done to clarify under what conditions subjects will be \textit{willing} and \textit{able} to contribute predictors.

We hypothesize that \textit{willingness} to generate candidate predictors of a behavioral outcome may be stimulated under several conditions. First, if subjects are incurring a health or financial cost as a result of the outcome under study, they may be motivated to contribute. For example a user that has an above average electricity bill or body mass index, yet has similar lifestyle attributes as his fellow users, may wish to generate additional attributes to explain the discrepancy.

Conversely, a user that posts a superior outcome (i.e. a low electricity bill or very healthy body mass index) may wish to uncover the predictor that contributes to their superior outcome (i.e. a well-insulated house or good exercise regimen) and thus advertise it to their peers. This may act as a form of online `boasting', a well known motivator among online communities.

In the current studies, some participants may have been motivated to contribute because they were part of the authors' social networks. However, a substantial number of users were recruited from online communities outside of the authors' social networks, indicating that some online users are motivated to contribute to such studies even if they do not know those responsible for the study. The exact number of users in these two groups is not clear on account of the anonymity requirements stipulated for these human subject studies.

Similarly, a non domain expert's \textit{ability} to contribute a previously unknown yet explanatory predictor of a behavioral outcome may rely on them suffering or benefiting from a far-from-average outcome. For example consider someone who is extremely underweight yet their outcome is not predicted by the common predictors of diet and exercise: this user has a high caloric intake and does not exercise. This user may be able to generate a predictor that a domain expert may not have though of, yet is predictive for a certain underweight demographic: this user may ask her peers: ``Are you in an abusive relationship?''

Users may also be motivated to contribute to such studies because it provides entertainment value: users may view the website as a competitive game in which the `goal' is to propose the best questions. In a future version we plan to create a dynamically sorted list of user-generated questions: questions bubble up to the top of the list if (1) it is a question that many other users wish to respond to, (2) it is orthogonal to the other questions, and (3) it is found to be predictive of the outcome under study. Users may then compete by generating questions that climb the leaderboard and thus advertise the user's understanding of the outcome under study.

\subsection{Rare Outcomes}

Obesity and electricity usage are well-studied behavioral outcomes. It remains to be seen though how the proposed methodology would work for outcomes that affect a small minority of online users, or for which predictors are not well known.

We hypothesize that for rare outcomes, online users who have experience with this outcome, could be encouraged to participate, and would be intrinsically motivated to contribute. For example if the outcome to be studied were a rare disease, users who suffer from the disease would be attracted to the site. Once there, they may be in a unique position to suggest and collectively discover previously unknown predictors of that disease. Moreover, a user who suffers from the disease is likely to know more people who suffer from that disease and would be motivated to advertise the site to them. Finally, even if a user discovers the site and does not suffer from the disease, he may know someone who does and thus introduce the site to that person. Such a person may serve as a caregiver for someone suffering from the disease, such as a family member. A caregiver may be able to contribute novel predictors that are different from those proposed by the sufferer himself.

Thus, a website that hosts such a rare outcome may serve as a `magnet' for people who exhibit the outcome or know people that do. In future we will study the 'attractive force' of such websites: if such a website experiences increased user traffic as the study goes forward, and the average outcome of users on the site drifts away from the global population's mean value for this outcome, that would indicate that a growing number of people with such an outcome are being attracted to the site.

In closing, this paper has presented a novel contribution to the growing
field of machine science in which the formulation of observables for
a modeling task---and the \emph{populating} of those observables with
values---can be offloaded to the human group being modeled.

\section{Acknowledgement}
The authors acknowledge valuable contributions from three anonymous reviewers, and useful discussions with collaborators in the UVM Complex Systems center.

\bibliographystyle{IEEEtran}
\bibliography{bmi-eminder}

\begin{IEEEbiography}[{\includegraphics[width=1in,height=1.25in,clip,keepaspectratio]{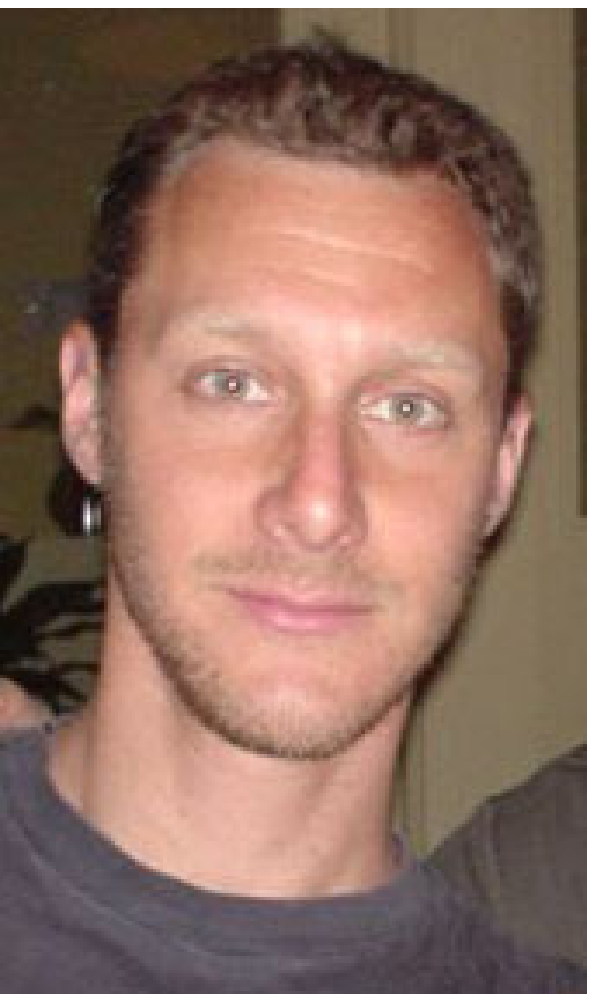}}]{Josh C. Bongard (M`06)}
is an Assistant Professor in the Department of Computer Science at the University of Vermont. Prior to this appointment, he was a postdoctoral researcher in the Sibley School of Mechanical and Aerospace Engineering at Cornell University. He received the B.Sc. honors degree in computer science from McMaster University, Canada, in 1997, and a M.Sc. in evolutionary and adaptive systems from the School of Cognitive and Computing Sciences at University of Sussex in 1999. He obtained his Ph.D. from the Artificial Intelligence Laboratory at the University of Zurich, for research in the field of evolutionary robotics. He was named a Microsoft New Faculty Fellow, one of the top 35 innovators under the age of 35 by MIT`s Technology Review Magazine, and is a National Science Foundation CAREER award recipient.
\end{IEEEbiography}

\begin{IEEEbiography}[{\includegraphics[width=1in,height=1.25in,clip,keepaspectratio]{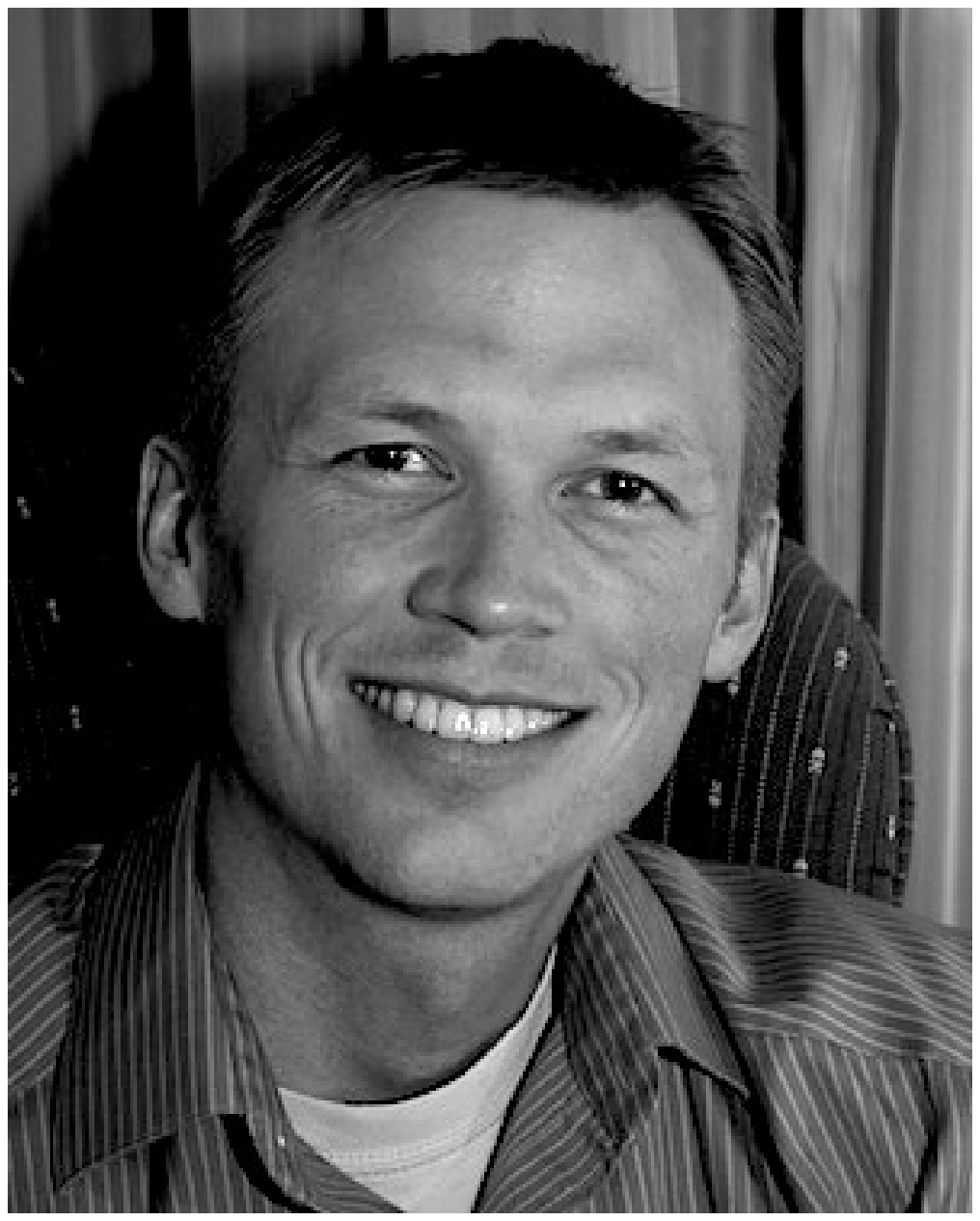}}]{Paul Hines (M`07)}
is an Assistant Professor in the School of Engineering at the University of Vermont. He is also a member of the Carnegie Mellon Electricity Industry Center Adjunct Research Faculty and a commissioner for the Burlington Electric Department. He received the Ph.D. in Engineering and Public Policy from Carnegie Mellon U. in 2007 and the M.S. degree in Electrical Engineering from the U. of Washington in 2001. Formerly he worked at the US National Energy Technology Laboratory, where he participated in Smart Grid research, the US Federal Energy Regulatory Commission, where he studied interactions between nuclear power plants and power grids, Alstom ESCA, where he developed load forecasting software, and for Black and Veatch, where he worked on substation design projects. His main research interests are in the areas of complex systems and networks, cascading failures in power systems, wind integration and energy security policy.
\end{IEEEbiography}

\begin{IEEEbiographynophoto}{Dylan Conger}
received the B.S. degree in Computer Science at the University of Vermont in 2011. He worked on the BMI project as an undergraduate research assistant in 2010.
\end{IEEEbiographynophoto}

\begin{IEEEbiographynophoto}{Peter Hurd}
received the M.S. degree in Computer Science from the University of Vermont in 2010. He worked on the EnergyMinder project as a graduate research assistant in 2009.
\end{IEEEbiographynophoto}

\begin{IEEEbiography}[{\includegraphics[width=1in,height=1.25in,clip,keepaspectratio]{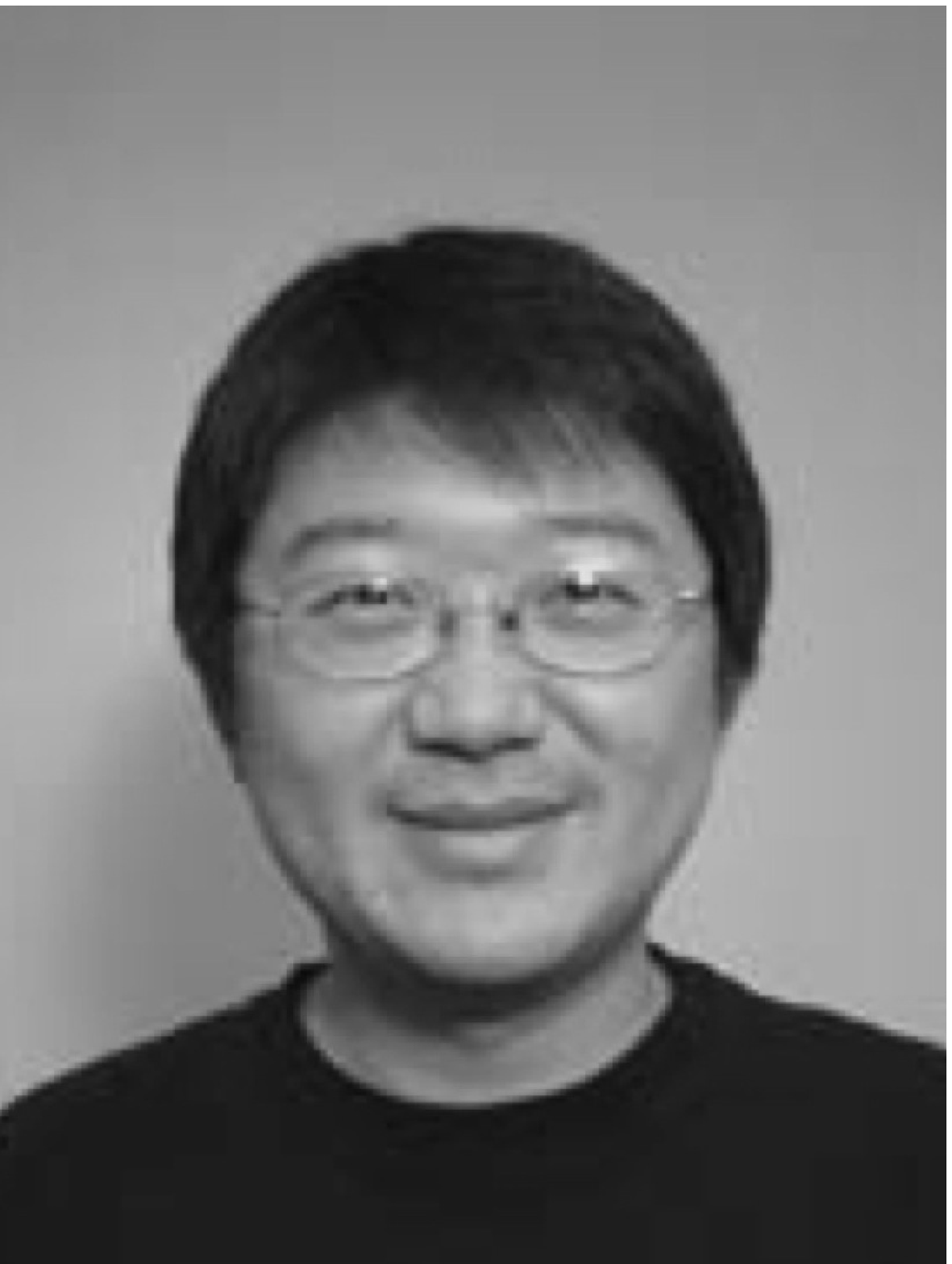}}]{Zhenyu Lu}
is a Ph.D candidate in Computer Science at the University of Vermont. His main research interests are active learning and ensemble learning. He has published in various data mining forums, including ACM SIGKDD, IEEE ICDM and SIAM SDM. He is currently working as a machine learning specialist at Sears Holdings Corporation (SHC) with focuses on Web information exploration.
\end{IEEEbiography}

\end{document}